\documentclass[aps,prd,reprint,amsmath,amssymb,superscriptaddress,floatfix]{revtex4-2}

\usepackage{graphicx}
\usepackage[colorlinks=true, allcolors=blue]{hyperref}

\usepackage{float}
\usepackage{dcolumn}

\usepackage[dvipsnames]{xcolor}
\usepackage[normalem]{ulem}

\newcommand{\replaceR}[2]{{{\color{BrickRed}{#1}}{\color{NavyBlue}{\ifmmode\text{\sout{\ensuremath{#2}}}\else\sout{#2}\fi}}}}

\newcommand{\replaceB}[2]{{{\color{RedViolet}{#1}}{\color{BlueViolet}{\ifmmode\text{\sout{\ensuremath{#2}}}\else\sout{#2}\fi}}}}

\newcommand{\replaceK}[2]{{{\color{RedOrange}{#1}}{\color{OliveGreen}{\ifmmode\text{\sout{\ensuremath{#2}}}\else\sout{#2}\fi}}}}

\begin{document}

\title{QCD Equation of State at nonzero baryon density in external magnetic field}

\author{N.~Astrakhantsev}
\email{nikita.astrakhantsev@physik.uzh.ch}
\affiliation{Department of Physics, University of Z\"urich, CH-8057 Z\"urich, Switzerland}
\author{V.~V.~Braguta}
\email{vvbraguta@theor.jinr.ru}
\affiliation{Bogoliubov Laboratory of Theoretical Physics, Joint Institute for Nuclear Research, 141980 Dubna, Russia}
\author{A.~Yu.~Kotov}
\email{akotov@theor.jinr.ru}
\affiliation{Bogoliubov Laboratory of Theoretical Physics, Joint Institute for Nuclear Research, 141980 Dubna, Russia}
\author{A.~A.~Roenko}
\email{roenko@theor.jinr.ru}
\affiliation{Bogoliubov Laboratory of Theoretical Physics, Joint Institute for Nuclear Research, 141980 Dubna, Russia}

\begin{abstract}
This paper is devoted to the study of QCD equation of state in external magnetic field and nonzero baryon density. Our study is carried out by means of lattice simulation with 2+1 dynamical staggered quarks at the physical masses. 
The simulation is conducted at imaginary baryon chemical potential what allowed us to overcome the sign problem. We expand the pressure in the baryon imaginary chemical potential and study three leading nonzero coefficients in this 
expansion. These coefficients were calculated for the following values of magnetic field: $eB=0.3$, $0.6$, $1.2$~GeV$^2$
with the lattice sizes $8\times32^3$, $10\times40^3$, $12\times48^3$.  Using these data 
we  take continuum limit for the coefficients. Our results indicate considerable enhancement of the expansion coefficients by the magnetic field. 
\end{abstract}

\maketitle

\section{Introduction}

Equation of State (EoS) of Quantum Chromodynamics (QCD) plays a fundamental role both from theoretical and practical points of view. From the theoretical perspective EoS contains an important information about thermodynamic properties and QCD phase transitions. On the other hand, from the practical  perspective EoS is used for hydrodynamic simulations of heavy-ion collision experiments as well as in different astrophysical applications. 

Despite lots of phenomenological works (see, for instance, \cite{Huovinen:2009yb, Andersen:2011sf, Soloveva:2021quj}), the most reliable information on QCD EoS can be  obtained by means of lattice QCD simulations. At zero baryon density it was studied in papers \cite{Borsanyi:2010cj, Borsanyi:2013bia, HotQCD:2014kol, Bernard:2006nj, Bazavov:2009zn}. Extension of lattice EoS studies to non-zero baryon chemical potential was conducted in papers \cite{Borsanyi:2012cr, Guenther:2017hnx, Bazavov:2017dus, DElia:2016jqh}. 

It is believed that quark-gluon matter created in non-central heavy-ion collisions is affected by strong magnetic field \cite{Kharzeev:2007jp, Skokov:2009qp}. For instance, it is expected that at RHIC magnetic field can reach the magnitude $eB \sim m_{\pi}^2$, while at LHC it is about $eB \sim 15 m_{\pi}^2$ \cite{Skokov:2009qp}. Such a strong magnetic field gives rise to multiple interesting phenomena. 
Probably the most famous example is the chiral magnetic effect which consists in the generation of electric current along magnetic field \cite{Fukushima:2008xe, Kharzeev:2013ffa}. In addition, strong magnetic field can noticeably modify QCD properties. In particular, one could mention the influence of magnetic field on QCD thermodynamics \cite{Buividovich:2008wf, DElia:2011koc, Bali:2012zg, Bruckmann:2013oba}, QCD phase diagram \cite{DElia:2010abb, Bali:2011qj, Ilgenfritz:2013ara, Ding:2020inp, Endrodi:2015oba, Braguta:2019yci, DElia:2021yvk}, the transport properties of quark gluon plasma \cite{Astrakhantsev:2019zkr} hadron spectroscopy \cite{Luschevskaya:2014lga, Bonati:2015dka, Luschevskaya:2015cko, Bali:2017ian, Ding:2020hxw} and etc.

Strong magnetic field might have significant impact on QCD EoS. This can be anticipated if one accounts for  dimensional reduction phenomenon realised in magnetic field \cite{Gusynin:1995nb}. The idea is that for small magnetic field the quarks 
live in $3+1$ dimensions, whereas for strong magnetic field they effectively live in $1+1$ dimensions. It is reasonable to expect that this phenomenon might lead to strong dependence of EoS on magnetic field. At zero baryon density EoS in external magnetic field was studied on the lattice in \cite{Bonati:2013vba, Levkova:2013qda, Bali:2014kia}. The second-order fluctuations of the baryon number, electric charge and strangeness, which are related to EoS, in external magnetic field were studied in papers \cite{Ding:2021cwv, MarquesValois:2023ehu}.
It is worth to note that perturbatively QCD EoS at nonzero density and magnetic field was studied in Refs.~\cite{Fraga:2023cef, Fraga:2023lzn}.

In this paper we are going to study EoS of QCD in external homogeneous magnetic field and nonzero baryon chemical potential. Our study is carried out within lattice simulation with dynamical staggered $u$-, $d$-, $s$-quarks at their physical masses. 
To avoid the sign problem the simulation is performed at imaginary chemical potentials and at the following setup  
$\mu_u=\mu_d=\mu,~\mu_s=0$ which approximately corresponds to zero strangeness.  We expand the free energy into a series in imaginary chemical potential and restrict our study to three nonzero terms in this expansion. Mainly we focus on the coefficients of this expansion and their dependence on magnetic field.  

This paper in organized as follows. In the next Section we derive the formulas important for our study. The lattice setup used in the simulation is presented in Section~3. In Section~4 the results of our study are shown. In last Section we discuss the results and draw  conclusions.

\section{Equation of State at finite baryon density in external magnetic field}

For a thermodynamic ensemble with a known partition function $Z$, the pressure $p$ may be found using the general formula
\begin{equation} \label{eq:eos1}
    \frac {p} {T^4} = \frac 1 {VT^3} \log {Z(V,T,\mu_u,\mu_d,\mu_s,eB )},
\end{equation}
where $\mu_u$, $\mu_d$, $\mu_s$ are the chemical potentials for $u$-, $d$- and $s$-quarks correspondingly. These chemical potentials can be rewritten in terms of the chemical potentials $\mu_B, \mu_Q, \mu_S$ which correspond to the baryon charge $B$, electric charge $Q$ and strangeness S.  The formulas relating the chemical potentials in different basis are well known
\begin{subequations}\label{eq:mu_syst}
\begin{align}
    \mu_u&=\mu_B/3+2\mu_Q/3\,, \\
    \mu_d&=\mu_B/3-\mu_Q/3\,,  \\
    \mu_s&=\mu_B/3-\mu_Q/3-\mu_S\,. 
\end{align}
\end{subequations}
In this paper we restrict our consideration to the following relation between quark chemical potentials: $\mu_u=\mu_d=\mu,\, \mu_s=0$, which approximately corresponds to zero strangeness $\langle S \rangle\simeq 0$. In this case Eqs.~\eqref{eq:mu_syst} give $\mu_B=3\mu,\,\mu_Q=0,\,\mu_S=\mu$. More accurate tuning of the chemical potential values relevant for modern heavy-ion experiments can be found in Refs.~\cite{Guenther:2017hnx, MarquesValois:2023ehu}. 

For a sufficiently small baryon density the EoS (\ref{eq:eos1}) can be expanded in a series in $\mu_B/T$:
\begin{multline}\label{eq:pressure}
    \frac{p(\mu_B)}{T^4}=c_0+c_2\left(\frac{\mu_B}{T}\right)^2+c_4\left(\frac{\mu_B}{T}\right)^4+{} \\ c_6\left(\frac{\mu_B}{T}\right)^6 + O\left(\left(\frac{\mu_B}{T}\right)^8\right)\,.
\end{multline}
The coefficient $c_0$ determines QCD EoS at zero baryon density and external magnetic field. This case was studied in papers \cite{Bonati:2013vba, Levkova:2013qda, Bali:2014kia} and we are not going to consider it. On contrary  in this paper we are going to focus on the coefficients $c_2,c_4, c_6$ and their dependence on the magnitude of external magnetic field. 

The coefficients $c_2,c_4, c_6$ are related to the fluctuations $\chi^{BQS}_{ijk}$ of the conserved charges $B,Q,S$ at zero baryon density 
\begin{equation}
\chi^{BQS}_{ijk}=\left.\frac{\partial^{i+j+k}(p/T^4)}{ \partial^i(\mu_B/T)\partial^j(\mu_Q/T)\partial^k(\mu_S/T)}\right|_{\mu_B=\mu_Q=\mu_S=0}.
\end{equation}
It causes no difficulties to find the formulas which connect the $\chi^{BQS}_{ijk}$ to the $c_2,c_4, c_6$ coefficients 
\begin{align}
c_2(&\mu_u=\mu_d=\mu,\mu_s=0)=\frac{1}{2!\,3^2}(9\chi_2^B+6\chi_{11}^{BS}+\chi_2^S)\,, \nonumber
    \\
c_4(&\mu_u=\mu_d=\mu,\mu_s=0)=\frac{1}{4!\,3^4}(81\chi_4^B+108\chi_{31}^{BS}+{} \nonumber \\ & \hspace{11em} 54\chi_{22}^{BS}+12\chi_{13}^{BS}+\chi_{4}^{S})\,,  \nonumber
    \\
c_6(&\mu_u=\mu_d=\mu,\mu_s=0)=\frac{1}{6!\,3^6}(729\chi_6^B+1458\chi_{51}^{BS}+{} \nonumber \\ & 1215\chi_{42}^{BS}+540\chi_{33}^{BS}+135\chi_{24}^{BS}+18\chi_{15}^{BS}+\chi_6^S)\,.
\label{eq:c246tochiBQS}
\end{align}
Lattice simulations at real values of the chemical potential are hampered by the sign problem \cite{Muroya:2003qs}, so we perform the simulations at the imaginary values of the chemical potential, characterized by $\theta=\mu_I/T=i\mu_B/T$ and expand our results in $\theta$. Using the data we can conduct an analytical continuation to the real values of $\mu$ as long as we do not encounter any discontinuity.
Since the thermal QCD phase transitions at baryon density and not too large magnetic field are expected to be a crossover, we believe that the analytical continuation procedure is justified in our case. 

The partition function itself and the pressure cannot be measured in the lattice simulations. One can only measure pressure derivatives with respect to external parameters. The basic quantity, which we measure on the lattice is the (imaginary) baryon density $n_I$, which is defined as the derivative of the pressure with respect to $\theta=i\mu_B/T$:
\begin{equation}
    \frac{n_I}{T^3}=\frac{\partial (p/T^4)}{\partial \theta}=-2 c_2\theta+4 c_4\theta^3-6 c_6\theta^5+O(\theta^7)\,.
    \label{eq:ni}
\end{equation}
By fitting the dependence of the imaginary baryon density $n_I$ on the values of the (imaginary) chemical potential $\theta$ we can extract the values of the coefficients $c_2$, $c_4$ and $c_6$.

\section{Lattice setup}

We perform simulations in QCD with $N_f=2+1$ staggered fermions at the physical pion mass in the presence of an external magnetic field $B$  and quark chemical potentials $\mu_f$ . The partition function $Z$ of the system under study has the following form:
\begin{multline}
    Z = \int DU_{x,\mu}\, e^{-S_g[U_{x,\mu}]}\left(\det D_u[U_{x,\mu},q_uB,\mu_u]\right)^{1/4} \times {} \\
    \det \left(D_d[U_{x,\mu},q_dB,\mu_d]\right)^{1/4} \times {} \\
    \det \left(D_s[U_{x,\mu},q_sB,\mu_s]\right)^{1/4},
\end{multline}
where $\det D_f[U_{x,\mu},q_fB,\mu_f]$ corresponds to the staggered Dirac operator of the quark flavour $f$. Note that various quark flavours have different quark charges $q_f$ and values of the chemical potential $\mu_f$, thus we treat all three flavours separately. Each staggered Dirac operator $\det D_f[U_{x,\mu},q_fB,\mu_f]$ corresponds to $4$ quark tastes and we take the fourth root to have one quark taste. Expression for the $D_f[U_{x,\mu},q_fB,\mu_f]$ has the following form:
\begin{multline}
    D_{f}[U_{x,\mu},q_fB,\mu_f]_{x,y}=am_f\delta_{x,y}+{} \\
    \frac12\sum_{\mu=1}^{4}\eta_{\mu}\Big(
    e^{a\mu_f\delta_{\mu,4}}u_{x,\mu}U^{(2)}_{x,\mu}\delta_{x,y-\mu}-{} \\
    e^{-a \mu_f\delta_{\mu,4}}u^*_{x-\mu,\mu}(U^{(2)}_{x-\mu,\mu})^{\dag}\delta_{x,y+\mu}
    \Big),
\end{multline}
where $am_f$ is the quark mass, $\eta_{\mu}=(-1)^{x_1+\ldots+x_{\mu-1}}$ are the standard factors for the staggered fermions, and $U^{(2)}_{x,\mu}$ are two-times stout smeared gauge links $U_{x,\mu}$ with isotropic smearing parameter $\rho=0.15$ \cite{Morningstar:2003gk}. The factors $u^f_{x,\mu}\in U(1)$ correspond to the magnetic field $B$  pointing in the $z$-direction and they are given by the following expressions ($N_s$ is the spatial lattice size): 
\begin{subequations}
\begin{alignat}{2}
 u^q_{x,1}&=e^{-ia^2q_fBx_2/2},        \qquad && x_1\ne N_s-1\,, \\
 u^q_{x,1}&=e^{-ia^2q_fB(N_s+1)x_2/2}, \qquad && x_1= N_s-1\,, \\
  u^q_{x,2}&=e^{ia^2q_fBx_1/2},        \qquad && x_2\ne N_s-1\,, \\
 u^q_{x,2}&=e^{ia^2q_fB(N_s+1)x_1/2},  \qquad &&  x_2= N_s-1\,, \\
 u^q_{x,3}&=u^q_{x,4}=1\,.     &&     
\end{alignat}
\end{subequations}

On the lattice with the periodic boundary conditions the magnetic flux is quantized. Since different quarks have different quarks charges, we take the minimum absolute value for the quark charge $q=|q_d|=|q_s|=e/3$ and the quantization condition for the magnetic field value takes the form
\begin{equation}
    eB=\frac{e}{q}\times qB=\frac{6\pi n}{N_s^2a^2}, \qquad n\in \mathbb{N}, \quad 0\le n < N_s^2\,.
\label{eq:quant}
\end{equation}

It should be noted that the condition $n\ll N_s^2$ is required in order to avoid lattice artifacts. In our simulations this condition is also fulfilled, moreover the comparison of results obtained at different values of $N_t$ suggest rather small lattice artifacts almost for all points used in the simulations.
Another restriction comes from the existence of the Roberge-Weiss phase transition \cite{Roberge:1986mm} in the plane of complex chemical potentials, $\mu_I/T\leq \pi$, which is also always fulfilled in our simulations. 

For the gluon sector we use the tree-level Symanzik improved gauge action:
\begin{equation}
    S_g = -\frac{\beta}{3}\sum_{x,\mu\ne\nu}\left(\frac{5}{6}W_{x,\mu\nu}^{1\times1}-\frac{1}{12}W_{x,\mu\nu}^{1\times2}\right),
\end{equation}
where $\beta={6}/{g^2}$ corresponds to the bare lattice gauge coupling and $W_{x,\mu\nu}^{n\times k} $ stands for the trace of the rectangular with the size $n\times k$ constructed from the gauge links $U_{x,\mu}$, starting from the point $x$ in directions $\mu$ and $\nu$.

Bare parameters have been set so as to stay on a line of constant
physics~\cite{Aoki:2009sc, Borsanyi:2010cj, Borsanyi:2013bia},
at the isospin symmetric point, $m_u=m_d=m_l$, a physical strange-to-light mass ratio, $m_s/m_l=28.15$, and a physical pseudo-Goldstone pion mass, $m_{\pi}\simeq135~\text{MeV}$.

We perform simulations with three lattice sizes $8\times32^3$, $10\times40^3$ and $12\times48^3$, keeping the aspect ratio $N_s/N_t=4$. Previous simulations \cite{Kolomoyets:2021onx} suggest that such aspect ratio has sufficiently small finite volume effects. By using three different lattice spacings we can study the continuum extrapolation of the studied quantities. Typically, the results for different $N_t$ are close to each other and we perform continuum extrapolation using simple quadratic ansatz for the dependence on the lattice spacing $a$: $O(a)=O(0)+Aa^2$. To assess the systematic uncertainty, we do the same procedure keeping only two fine lattice sizes $N_t=10$ and $N_t=12$ and take half of the sum as our final result and half of the difference as the systematic uncertainty.

\section{Results of the calculation and discussion}

%
\begin{figure}
    \centering
    \includegraphics[width=0.99\linewidth]{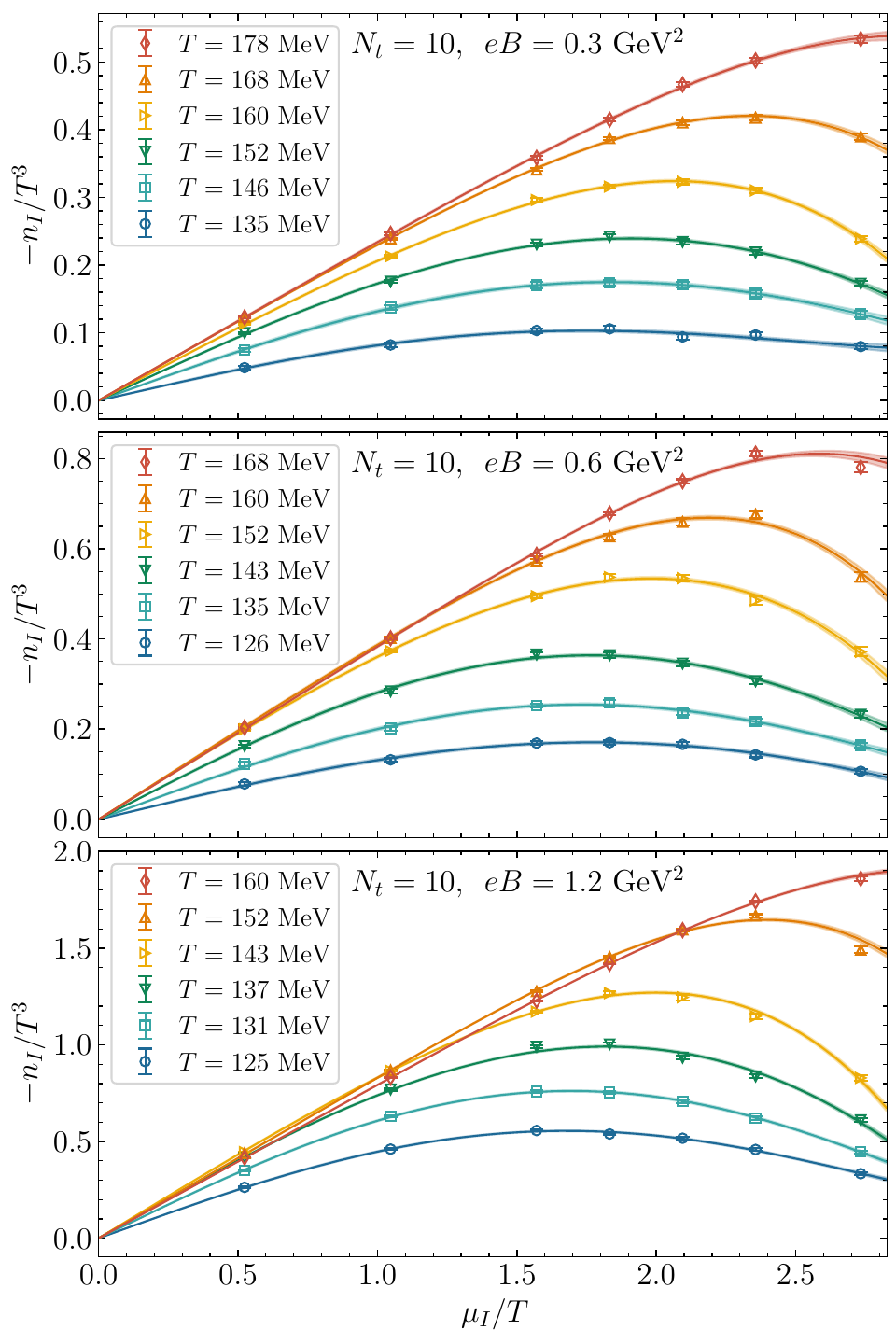}
    \caption{The density $n_I/T^3$ as a function of the imaginary chemical potential $\mu_I/ T=\theta$ calculated on the lattice $10\times40^3$ for three values of the magnetic field $eB=0.3,~0.6,~1.2$ GeV$^2$ and several temperatures in the vicinity of the thermal phase transition.}
    \label{fig:nvstheta}
\end{figure}

%
\begin{figure*}
    \centering
    \includegraphics[width=0.320\linewidth]{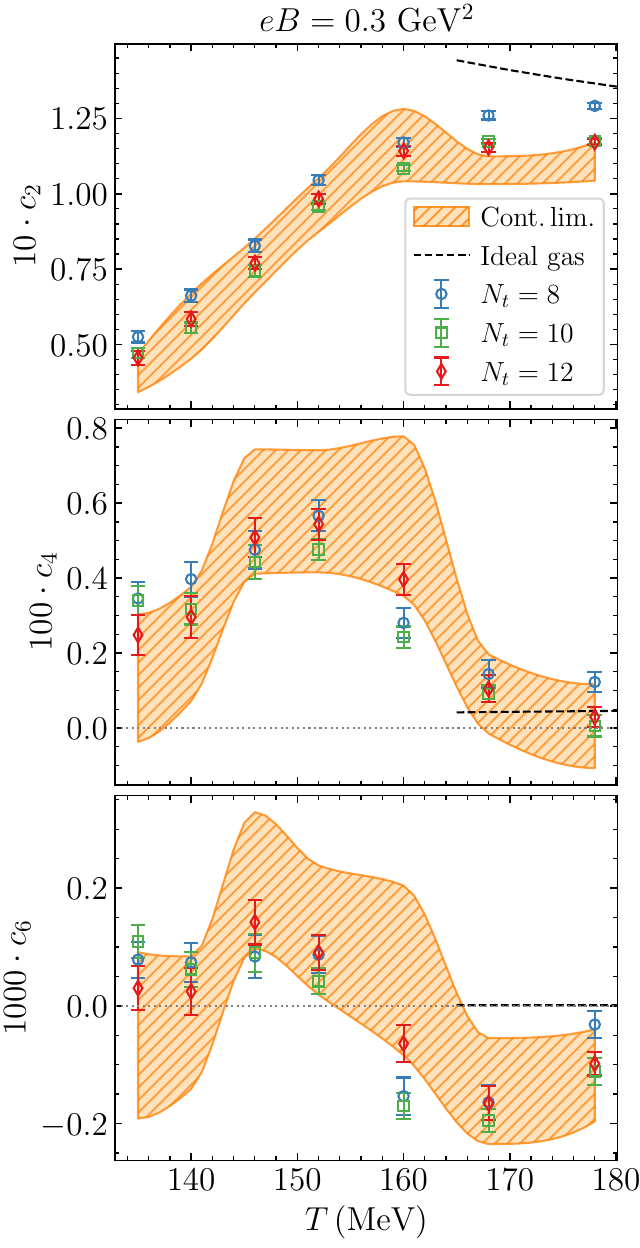}
    \includegraphics[width=0.320\linewidth]{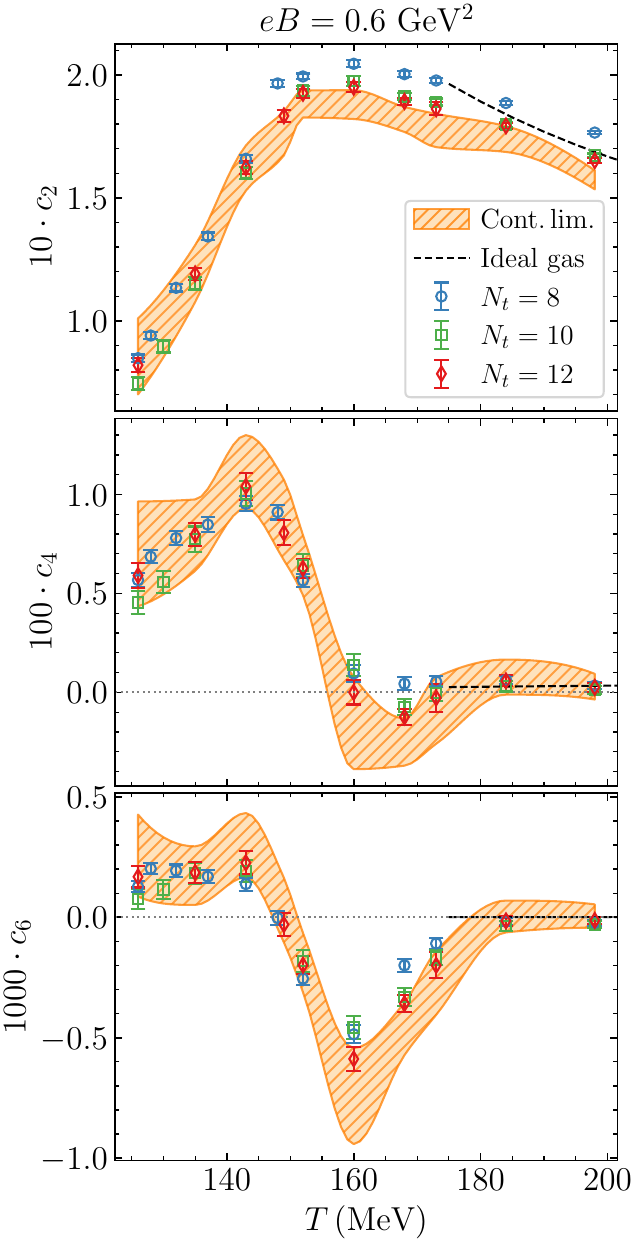}
    \includegraphics[width=0.320\linewidth]{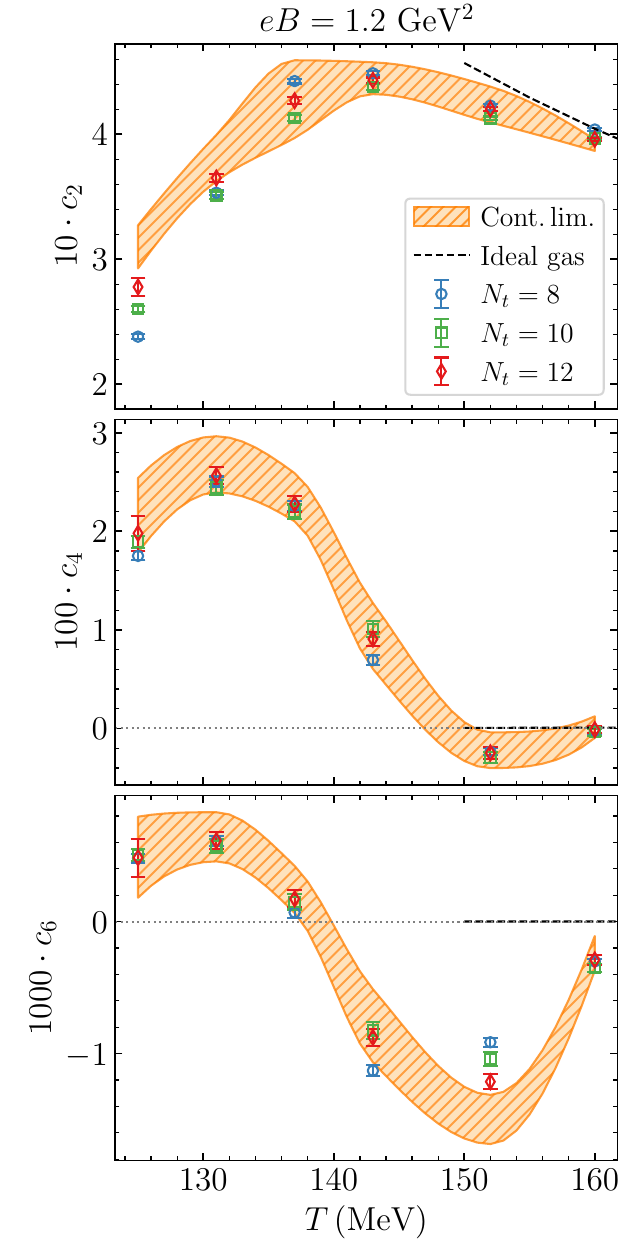}
    \caption{The coefficients $c_2$, $c_4$, $c_6$ 
    as functions of temperature for all studied values of magnetic field calculated on the lattices: $8\times32^3$, $10\times40^3$, $12\times48^3$. The orange bands correspond to the continuum extrapolation of the coefficients, their widths reflect full uncertainties, combined from statistical and systematic. The dashed line represents the coefficients in the ideal gas approximation.}
    \label{fig:c246_allb}
\end{figure*}

%
\begin{figure}
    \centering
    \includegraphics[width=0.99\linewidth]{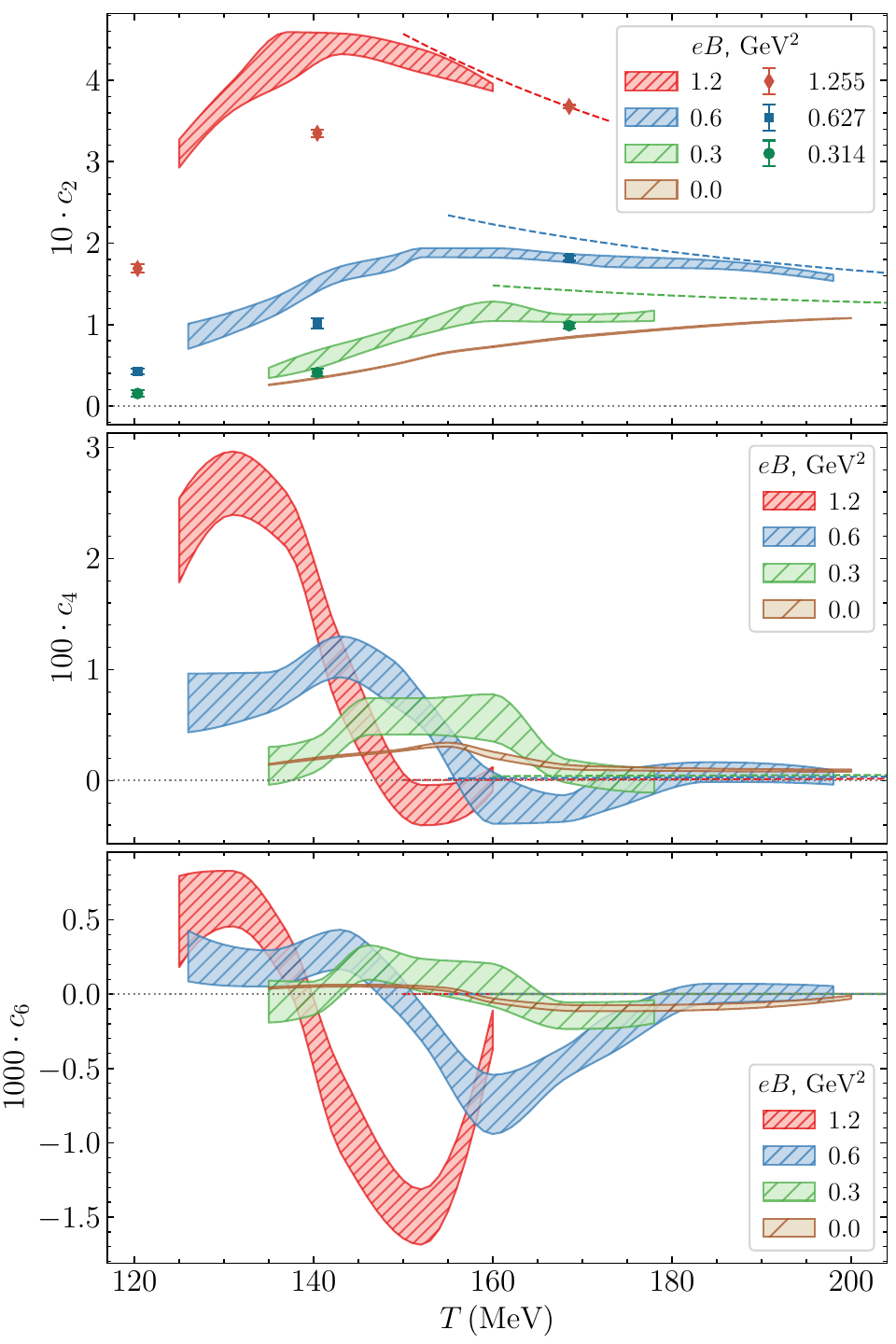}
    \caption{
Continuum extrapolated coefficients $c_2$, $c_4$, $c_6$ as  functions of temperature. The dashed lines represent the ideal gas approximation for the same coefficients. 
Various magnetic fields considered in this work are marked by colors as follows $eB=0.3$~GeV$^2$~(green),~$0.6$~GeV$^2$~(blue),~$1.2$~GeV$^2$~(red). In addition we show the results for the $c_2$ coefficient \cite{Ding:2021cwv} obtained at the magnetic fields $eB=1.255$~GeV$^2$ (red diamonds), $eB=0.627$~GeV$^2$ (blue squares), $eB=0.314$~GeV$^2$ (green circles) and the coefficients $c_2$, $c_4$, $c_6$  at zero magnetic field (brown bands) determined from \cite{DElia:2016jqh}.
    }
    \label{fig:c246_all}
\end{figure}

%
\begin{figure*}
    \centering
    \includegraphics[width=0.320\linewidth]{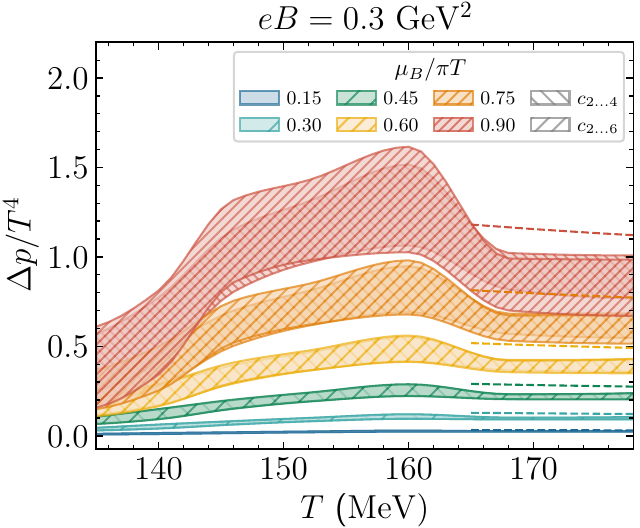}
    \includegraphics[width=0.320\linewidth]{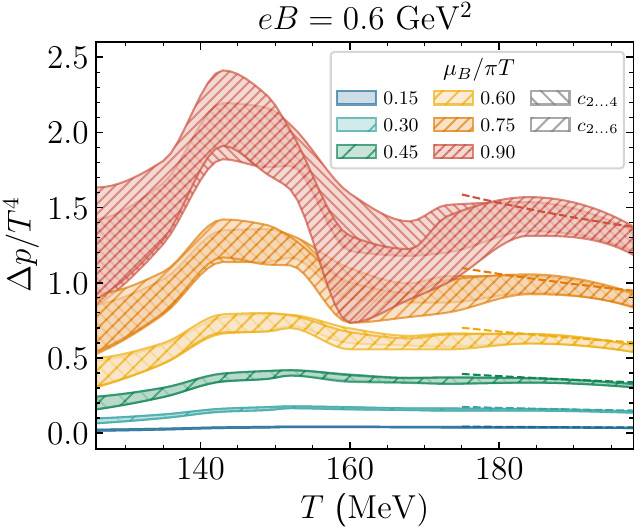}
    \includegraphics[width=0.320\linewidth]{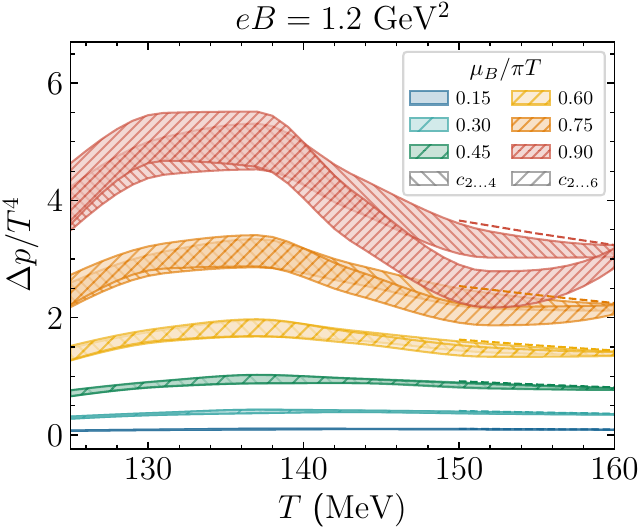}
    \caption{
The excess of pressure $\Delta p = p(\mu_B) - p(\mu_B=0)$ due to nonzero baryon density, reconstructed using Eq.~\eqref{eq:pressure}, as a function of temperature for the values of (real) baryon chemical potential $\mu_B/\pi T = 0.15,\, 0.30,\, 0.45,\, 0.60,\, 0.75,\, 0.90$ for all magnetic fields used in our study.  
Different directions of diagonal hatching correspond to the results obtained using all three coefficients $c_2,\, c_4,\, c_6$ or using only two lower coefficients $c_2,\, c_4$ in Eq.~\eqref{eq:pressure}.
The dashed lines represent the ideal gas limit.
 }
    \label{fig:press_all}
\end{figure*}

In Fig.~\ref{fig:nvstheta} we present the dependence of the density $n_I$ on the (imaginary) chemical potential $\mu_I$ for all three studied values of magnetic field $eB=0.3,\, 0.6$ and $1.2$~GeV$^2$ and for various temperatures in the vicinity of the phase transition. The data shown in Fig.~\ref{fig:nvstheta} were obtained on the lattice $10\times40^3$. The density $n_I$ for various $\mu_I$, $eB$ and $T$ was also calculated on the lattices $8\times32^3$, $12\times48^3$. 

Fitting our lattice data for the $n_I$ by formula (\ref{eq:ni}) we obtained the coefficients $c_2$,  $c_4$, $c_6$.
In Fig.~\ref{fig:c246_allb} we plot these coefficients as function of temperature for all values of magnetic field and lattices considered in our study. In addition in Fig.~\ref{fig:c246_allb} we show continuum limit for the coefficients $c_2$, $c_4$, $c_6$. 
The numerical values of these coefficients in continuum limit are listed in the Appendix~\ref{app:data}, and the bands in Fig.~\ref{fig:c246_allb} represent the interpolation of these data by the cubic splines~\cite{Fritsch1984}.

In Fig.~\ref{fig:c246_all} we show the continuum extrapolation for the  coefficients  $c_2$,  $c_4$, $c_6$, but contrary to Fig.~\ref{fig:c246_allb} we draw each coefficient for all values of magnetic field on one plot. 
In addition in Fig.~\ref{fig:c246_all} we present the results from Ref  \cite{Ding:2021cwv} obtained at the following values of magnetic field $eB=0.314,\, 0.627,\, 1.255$~GeV$^2$ and at zero magnetic field, determined from \cite{DElia:2016jqh}. 
In these works the authors calculated the fluctuations of the conserved charges, from which one can easily reconstruct the coefficients $c_2$, $c_4$ and $c_6$ using formulae (\ref{eq:c246tochiBQS}). In particular, the results of \cite{Ding:2021cwv} allow to determine the $c_2$ only, while using the results of \cite{DElia:2016jqh} one can find all three coefficients $c_2$, $c_4$, $c_6$ considered in this paper.
The results of \cite{Ding:2021cwv} were obtained at slightly larger pion mass $m_{\pi}\approx 220$ MeV and magnetic fields $eB=0.314,\, 0.627,\, 1.255$~GeV$^2$, which are very close to the values used in this work. We see good agreement with our results for the $c_2$ at high temperatures and some difference between them at $T\leq 140$~MeV. We believe, that this discrepancy can be attributed to the higher pion mass in \cite{Ding:2021cwv}. It might also be attributed to the fact that the authors of \cite{Ding:2021cwv} did not take continuum limit. 

Now few comments are in order.  First let us consider continuum extrapolated $c_2$ coefficient.  It seen in Fig.~\ref{fig:c246_all} that the $c_2$ coefficient, i.e. fluctuations, gradually increases with magnetic field. At the largest value of magnetic field $eB=1.2$~GeV$^2$ the $c_2$ is by an order of magnitude larger than that at zero magnetic field. Moreover magnetic field changes the dependence of the $c_2$ on temperature. Thus at zero magnetic field the $c_2$ coefficient is monotonously rising function of temperature, while for large magnetic fields considered in this paper the $c_2$ clearly develops a peak structure. For larger magnetic fields the peak is more pronounced.  In addition the position of the peak shifts to lower temperatures with increasing of the value of $eB$. We believe that this peak is related to the QCD critical temperature and the shift of the peak position to the left is related to inverse magnetic catalysis phenomenon~\cite{Bruckmann:2013oba}.  

Further let us pay attention to the $c_4$ and $c_6$ coefficients. They are known to exhibit nontrivial behaviour around the phase transition/crossover. For example, if one looks at the plot for $c_4$ (middle panel of Fig.~\ref{fig:c246_all}), then one observes a peak around the transition temperature. At the same time, the dependence of $c_6$ (lower panel of Fig.~\ref{fig:c246_all}) is even more complicated, it has positive peak before the transition and a negative dip after the phase transition. It can clearly seen, that corresponding peaks and dips are shifted to the left with increasing magnetic field, which is in agreement with inverse magnetic catalysis phenomenon. Moreover, their magnitude significantly grows with the magnetic field, while their width decreases.

Using our results for coefficients $c_2$, $c_4$ and $c_6$ we calculated additional contribution to the pressure coming from nonzero baryon density $\Delta p = p(\mu_B) - p(\mu_B=0)$. In order to perform it we carried out analytical continuation from imaginary to real values of the baryon chemical potential using formula (\ref{eq:pressure}). We would like to note here that the problem of analytical continuation is not well defined and becomes unstable at large values of baryon chemical potential. For this reason we consider values of chemical potential $\mu_B/\pi T \leq 0.9$, within this region the results from analytical continuation are stable against various functional forms for zero magnetic field \cite{Guenther:2017hnx}. In Fig.~\ref{fig:press_all} we plot the pressure excess $\Delta p$ as a function of temperature for various values of real baryon chemical potential $\mu_B$ and all magnetic fields considered in this work. To assess the stability of our results, we determined the pressure excess $\Delta p$ either using all three coefficients $c_2$, $c_4$ and $c_6$ or using only two lower coefficients $c_2$ and $c_4$ (shown in Fig.~\ref{fig:press_all} by different hatching). It can be clearly seen that the results obtained in two different ways are almost the same, only for the largest studied value of chemical potential $\mu_B/\pi T=0.90$ there is some difference. It confirms that for the studied values of $\mu_B$ the analytical continuation is under control. 

From the Fig.~\ref{fig:press_all} one can conclude that magnetic field not only enhances the pressure $\Delta p$ but also modifies its dependence on temperature and chemical potential. 

It is believed that at high temperatures the quark-gluon plasma reveal properties of ideal gas of quarks and gluons. In the ideal gas approximation magnetic field acts only on quarks. The pressure for massless quarks in magnetic field can be written  in the following form~\cite{Ding:2020inp} 
\begin{align}
    \frac {p} {T^4} &{}= \frac {8\pi^2} {45} + \sum_{f=u,d,s} 
    \frac {3 |q_f|B} {\pi^2 T^2} \biggl [ 
    \frac  {\pi^2} {12} + \frac {\hat \mu_f^2} {4} + p_f(B)
    \biggr ]\,,  \nonumber
    \\
    p_f(B) &{}= 2 \frac {\sqrt{2 |q_f| B}} {T} \sum_{l=1}^{\infty} \sqrt{l} \sum_{k=1}^{\infty} 
    \frac {(-1)^{k+1}} {k} \cosh (k \hat \mu_f) \times {} \nonumber \\
    & \hspace{10em} K_1 \biggl ( \frac {k \sqrt {2 |q_f| B l}} {T} \biggr ),
    \label{eq:eos_gas}
\end{align}
where we have used the designation $\hat \mu_f = {\mu_f} / {T}$.
Expression (\ref{eq:eos_gas}) allows us to find the coefficients $c_2$, $c_4$, $c_6$  for an ideal gas which have the following form
\begin{align}
\label{eq:idela_gas_c2}
 c_2 ={} & \frac 1 {12} \frac {eB} {\pi^2 T^2} + \frac 1 {3} \frac {eB} {\pi^2 T^2} \sum_{f=u,d} \frac{|q_f|}{e}  \frac {\sqrt{2 |q_f| B}} {T} \times {} \nonumber \\
 & \quad \sum_{l=1}^{\infty} \sqrt{l} \sum_{k=1}^{\infty} {(-1)^{k+1}} k\,  K_1 \biggl ( \frac {k \sqrt {2 |q_f| B l}} {T} \biggr )\,, \\
 \label{eq:idela_gas_c4}
 c_4 ={} & \frac 2 {4!\, 3^3} \frac {eB} {\pi^2 T^2} \sum_{f=u,d} \frac{|q_f|}{e}  \frac {\sqrt{2 |q_f| B}} {T} \times {} \nonumber \\
 & \quad \sum_{l=1}^{\infty} \sqrt{l} \sum_{k=1}^{\infty}  {(-1)^{k+1}} k^3 K_1 \biggl ( \frac {k \sqrt {2 |q_f| B l}} {T} \biggr )\,, \\
 \label{eq:idela_gas_c6}
 c_6 ={} & \frac 2 {6!\, 3^5} \frac {eB} {\pi^2 T^2} \sum_{f=u,d} \frac{|q_f|}{e}  \frac {\sqrt{2 |q_f| B}} {T} \times {} \nonumber \\
 & \quad \sum_{l=1}^{\infty} \sqrt{l} \sum_{k=1}^{\infty}  {(-1)^{k+1}} k^5  K_1 \biggl ( \frac {k \sqrt {2 |q_f| B l}} {T} \biggr )\,.
\end{align}
It should be noted here that the first term in the coefficient $c_2$  (\ref{eq:idela_gas_c2}) arises from the lowest Landau level, whereas the second term results from the excited Landau levels. What concerns the coefficients $c_4$, $c_6$ (\ref{eq:idela_gas_c4}), (\ref{eq:idela_gas_c6}) they acquire contribution from the excited Landau levels only.  The values of magnetic fields studied in this paper obey the inequality  $eB > T^2$. Moreover, for the largest magnetic field it takes the form of $eB \gg T^2$.  So, in the ideal gas approximation the contribution of the exited Landau levels are suppressed leading to the suppression of the coefficients $c_4$ and $c_6$.  However, radiative corrections to these coefficients might change this picture. 

In Fig.~\ref{fig:c246_allb} and Fig.~\ref{fig:c246_all}
we also show the coefficients  $c_2$, $c_4$, $c_6$ in the ideal gas approximation. First let us consider the $c_2$ coefficient. Our continuum limit for the $c_2$ and ideal gas approximation prediction differ by 20-30\% in the region $T>160$~MeV for the smallest magnetic field.  However, as one increases the  value of magnetic field this discrepancy decreases and at the largest magnetic field lattice results are in agreement with the ideal gas approximation. Taking into account the above discussion this result implies that the lowest Landau level plays more and more important role as one increases the magnitude of magnetic field in the temperature range $T\geq 160$~MeV. Probably for the largest magnetic field $eB=1.2$~GeV$^2$ the contribution of the lowest Landau level becomes dominant. What concerns the coefficients $c_4$, $c_6$  lattice results are in agreement with the ideal gas approximation in the temperature interval $T>160$~MeV. Notice, however, that the uncertainties of our lattice results are quite large in this case. 

In Fig.~\ref{fig:press_all} we have shown the excess of pressure $\Delta p$ in the ideal gas approximation by the dashed lines.  Similarly to  Fig.~\ref{fig:c246_allb} and Fig.~\ref{fig:c246_all}
lattice results for the $\Delta p$  are in agreement with the ideal gas approximation in the temperature interval $T>160$~MeV.

\section{Conclusion}

In this paper we conducted the study of QCD equation of state in external magnetic field and nonzero baryon density. 
Our study is carried out within lattice simulation with 2+1 dynamical staggered quarks at their physical masses. 
Nonzero baryon density was introduced to the system through baryon chemical potential. 
To avoid the sign problem the simulation is performed at imaginary chemical potential.  
We expand the pressure into a series in imaginary chemical potential and focus in our study on the first three  nonzero coefficients in this 
expansion. These coefficients were calculated on the lattices $8\times32^3$, $10\times40^3$, $12\times48^3$ for the following values of magnetic field: $eB=0.3$, $0.6$, $1.2$~GeV$^2$, and the continuum limit was taken. 

Our data indicate that magnetic field modifies the temperature dependence of the coefficients. For instance, large magnetic field gives rise to the peak of the $c_2$ coefficients which was absent at zero magnetic field. Notice also that for large magnetic fields positions of peaks and dips in the temperature dependence of coefficients are shifted to lower temperatures. It is reasonable to assume that nontrivial behaviour of the coefficients takes place in the crossover region and the shift of this region to lower temperatures is related to the inverse magnetic catalysis phenomenon.  

In addition to modifying the temperature dependence, magnetic field enhances the magnitude of the coefficients $c_2$, $c_4$, $c_6$ considerably. The coefficients are known to be connected to the fluctuations of the conserved charges, i.e. magnetic field enhances these fluctuations. The origin of this enhancement is not completely clear. It is might be related to the dimensional reduction phenomenon. 
Indeed, we observe considerable enhancement of the $c_2$ coefficient in an ideal fermion gas approximation, but this approximation is not sufficient to explain the behaviour of the $c_4$ and $c_6$ coefficients, possibly indicating that the enhancement of the fluctuations in the $c_4$ and $c_6$ coefficients is of nonperturbative nature.

\begin{acknowledgments}
The authors are grateful to Natalia Kolomoyets for the help at the initial stage of the project. This work has been carried out using computing resources of the Federal collective usage center Complex for Simulation and Data Processing for Mega-science Facilities at NRC ``Kurchatov Institute'', http://ckp.nrcki.ru/ and the Supercomputer ``Govorun'' of Joint Institute for Nuclear Research. This work was supported by the Russian Science Foundation (project no. 23-12-00072).
\end{acknowledgments}

\appendix*

\section{Data tables} \label{app:data}
The continuum limit values of the coefficients $c_2, c_4, c_6$ in the pressure expansion~\eqref{eq:pressure} at fixed magnetic fields $eB = 0.3,\, 0.6,\, 1.2$~GeV$^2$ are listed in Tables~\ref{tab:eB03},~\ref{tab:eB06},~\ref{tab:eB12}. 
Data are presented with full uncertainty, which includes both statistical and systematic errors. 

\begin{table}[H]
\caption{Continuum limit results for the coefficients of the series, Eq.~\eqref{eq:pressure}, at $eB = 0.3$~GeV$^2$.
}\label{tab:eB03}
\begin{ruledtabular}
\begin{tabular}{c D{.}{.}{8} D{.}{.}{8} D{.}{.}{8} }
\multicolumn{1}{c}{$T$,~\textrm{MeV}} &
\multicolumn{1}{c}{$10\cdot c_2 $ } &
\multicolumn{1}{c}{$100\cdot c_4 $} & 
\multicolumn{1}{c}{$1000\cdot c_6$ }\\
\colrule\noalign{\smallskip}
135 & 0.405(63) & 0.133(170) & -0.050(141) \\
140 & 0.561(108) & 0.224(152) & -0.029(113) \\
146 & 0.762(89) & 0.578(166) & 0.213(116) \\
152 & 0.963(89) & 0.578(163) & 0.129(110) \\
160 & 1.162(120) & 0.566(213) & 0.060(144) \\
168 & 1.078(46) & 0.091(106) & -0.145(90) \\
178 & 1.107(64) & 0.005(112) & -0.118(78) \\
\end{tabular}
\end{ruledtabular}
\end{table}

\begin{table}[H]
\caption{Continuum limit results for the coefficients of the series, Eq.~\eqref{eq:pressure}, at $eB = 0.6$~GeV$^2$.}\label{tab:eB06}
\begin{ruledtabular}
\begin{tabular}{c D{.}{.}{8} D{.}{.}{8} D{.}{.}{8} }
\multicolumn{1}{c}{$T$,~\textrm{MeV}} &
\multicolumn{1}{c}{$10\cdot c_2 $ } &
\multicolumn{1}{c}{$100\cdot c_4 $} & 
\multicolumn{1}{c}{$1000\cdot c_6$ }\\
\colrule\noalign{\smallskip}
126 & 0.855(155) & 0.700(265) & 0.256(170) \\
135 & 1.192(116) & 0.794(180) & 0.173(122) \\
143 & 1.621(92) & 1.115(184) & 0.299(133) \\
149 & 1.762(90) & 0.871(204) & 0.029(134) \\
152 & 1.882(55) & 0.643(152) & -0.188(123) \\
160 & 1.879(58) & -0.156(232) & -0.742(199) \\
168 & 1.819(51) & -0.251(119) & -0.464(109) \\
173 & 1.774(68) & -0.092(168) & -0.273(132) \\
184 & 1.739(52) & 0.077(89) & 0.003(66) \\
198 & 1.573(39) & 0.029(65) & 0.005(48) \\
\end{tabular}
\end{ruledtabular}
\end{table}

\begin{table}[H]
\caption{Continuum limit results for the coefficients of the series, Eq.~\eqref{eq:pressure}, at $eB = 1.2$~GeV$^2$.}\label{tab:eB12}
\begin{ruledtabular}
\begin{tabular}{c D{.}{.}{8} D{.}{.}{8} D{.}{.}{8} }
\multicolumn{1}{c}{$T$,~\textrm{MeV}} &
\multicolumn{1}{c}{$10\cdot c_2 $ } &
\multicolumn{1}{c}{$100\cdot c_4 $} & 
\multicolumn{1}{c}{$1000\cdot c_6$ }\\
\colrule\noalign{\smallskip}
125 & 3.099(172) & 2.163(378) & 0.488(307) \\
131 & 3.809(185) & 2.679(285) & 0.642(186) \\
137 & 4.280(311) & 2.345(245) & 0.244(178) \\
143 & 4.448(127) & 0.938(334) & -0.789(277) \\
152 & 4.237(143) & -0.221(181) & -1.499(186) \\
160 & 3.912(45) & 0.013(109) & -0.241(131) \\
\end{tabular}
\end{ruledtabular}
\end{table}

\bibliographystyle{apsrev4-2}
\bibliography{eosmagnetic}

\end{document}